\documentclass[sigconf, review]{acmart}
\renewcommand\footnotetextcopyrightpermission[1]{} % removes footnote with conference information in first column
\usepackage{xspace}
\usepackage{graphicx}
\newcommand\myname{PNR-LLM\xspace}

\usepackage{amsmath}
\usepackage{multirow}

\usepackage{xcolor}
\usepackage{subcaption}
\usepackage{listings}
\usepackage[normalem]{ulem}

\usepackage{graphicx}
\usepackage{wrapfig}
\usepackage{amsmath}
\usepackage{bm}
\usepackage{caption}
\usepackage{multirow}
\usepackage{mathrsfs}
\usepackage{mathtools}
\usepackage{algorithm}
\usepackage{algpseudocode}
\usepackage{soul}
\usepackage{booktabs}
\AtBeginDocument{%
  }

\acmISBN{978-1-4503-XXXX-X/18/06}

\begin{document}

\title{Enhancing News Recommendation with Hierarchical LLM Prompting}
% \title{Beyond the Title: Hierarchical Prompting is Revolutionizing News Recommendations}

\author{Hai-Dang Kieu}
\affiliation{%
  \institution{VinUniversity}
  \city{HaNoi}
  \country{VietNam}
}
\email{dang.kh@vinuni.edu.vn}

\author{Delvin Ce Zhang}
\affiliation{%
  \institution{The Pennsylvania State University }
  \city{State College}
  \country{United States}
}
\email{delvincezhang@gmail.com}

\author{Minh Duc Nguyen}
\affiliation{%
  \institution{VinUniversity}
  \city{HaNoi}
  \country{VietNam}
}
\email{duc.nm2@vinuni.edu.vn}

\author{Qiang Wu}
\affiliation{%
  \institution{University of Technology Sydney}
  \city{Sydney}
  \country{Australia}
}
\email{Qiang.Wu@uts.edu.au}

\author{Min Xu}
\affiliation{%
  \institution{University of Technology Sydney}
  \city{Sydney}
  \country{Australia}
}
\email{min.xu@uts.edu.au}

\author{Dung D. Le}
\affiliation{%
  \institution{VinUniversity}
  \city{HaNoi}
  \country{VietNam}
}
\email{dung.ld@vinuni.edu.vn}

\renewcommand{\shortauthors}{Dang Kieu et al.}

\begin{abstract}

Personalized news recommendation systems often struggle to effectively capture the complexity of user preferences, as they rely heavily on shallow representations, such as article titles and abstracts. To address this problem, we introduce a novel method, namely \myname, for \underline{\textbf{L}}arge \underline{\textbf{L}}anguage \underline{\textbf{M}}odels for \underline{\textbf{P}}ersonalized \underline{\textbf{N}}ews \underline{\textbf{R}}ecommendation. Specifically, \myname harnesses the generation capabilities of LLMs to enrich news titles and abstracts, and consequently improves recommendation quality. \myname contains a novel module, \textit{News Enrichment via LLMs}, which generates deeper semantic information and relevant entities from articles, transforming shallow contents into richer representations. We further propose an attention mechanism to aggregate enriched semantic- and entity-level data, forming unified user and news embeddings that reveal a more accurate user-news match. Extensive experiments on MIND datasets show that \myname outperforms state-of-the-art baselines. Moreover, the proposed data enrichment module is model-agnostic, and we empirically show that applying our proposed module to multiple existing models can further improve their performance, verifying the advantage of our design.
%Extensive experiments on the MIND datasets show that \myname not only consistently outperforms state-of-the-art methods but also empowers the performance of basic baseline models.

\end{abstract}

%    <concept>
%        <concept_id>10002951.10003317.10003338</concept_id>
%        <concept_desc>Information systems~Retrieval models and ranking</concept_desc>
%        <concept_significance>500</concept_significance>
%        </concept>
%    <concept>
%        <concept_id>10002951.10003317.10003338.10003341</concept_id>
%        <concept_desc>Information systems~Language models</concept_desc>
%        <concept_significance>500</concept_significance>
%        </concept>
%    <concept>
%        <concept_id>10002951.10003317.10003338.10003343</concept_id>
%        <concept_desc>Information systems~Learning to rank</concept_desc>
%        <concept_significance>500</concept_significance>
%        </concept>
%  </ccs2012>

% \ccsdesc[500]{Information systems~Retrieval models and ranking}
% \ccsdesc[500]{Information systems~Language models}
% \ccsdesc[500]{Information systems~Learning to rank}


\maketitle
\pagestyle{plain} % removes running headers
\section{Introduction}

With the rapid growth of online content, recommender systems play a crucial role in delivering personalized experiences to users. In the domain of news recommendation, the goal is to predict and recommend articles that align with individual users' interests. Given the fast-paced and dynamic nature of news, developing effective news recommender systems remains a challenging and important research problem \cite{ChuhanWuSurvey, LiuQijiongSurvey}.
% Content-based methods have proven effective in news recommendation by leveraging natural language processing and machine learning to model user preferences. These approaches analyze users' reading histories and construct news representations based on article content. CNNs and GRUs are widely used, as seen in FIM~\cite{FIM2020} and NAML~\cite{NAML}, while NRMS~\cite{NRMS} employs multi-head self-attention for dynamic modeling. MINER~\cite{miner} enhances user representations with poly-attention and category-aware weighting, and LSTUR~\cite{lstur} incorporates ID embeddings to capture long-term preferences.  

News recommendation systems face several challenges due to the unique characteristics of news articles and user behaviors. Moreover, news articles often contain sparse or ambiguous textual information, especially in short titles, which can limit accurate user preference modeling. While content-based methods focus on article text and reading histories but fail to capture structured relationships between articles. Graph-based methods leverage external knowledge graphs but suffer from sparsely connected or incomplete data, limiting their effectiveness.

\begin{figure}[ht!]
  \centering
  \includegraphics[width=.65\linewidth]{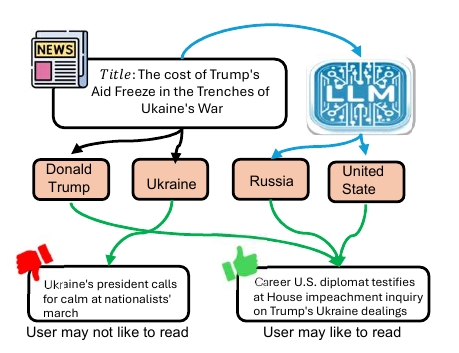}
  \label{fig:example}
  \vspace{-0.1in}
  \caption{Illustration of news connected to a relevant news through entities explored by LLM.}
  \vspace{-0.3in}
  
\end{figure}
% Beyond text-based methods, graph-based approaches improve recommendations by incorporating structured knowledge. DKN~\cite{DKN} integrates entity embeddings from knowledge graphs, and DIGAT~\cite{digat} introduces dual-interactive graph attention networks. GLORY~\cite{GLORY} refines user representations through graph-based entity extraction.  Despite their strengths, these methods face challenges. Long or unstructured documents introduce noise, diluting feature relevance, while short and vague titles fail to capture user intent. Additionally, content-based methods often overlook structured relationships, limiting their ability to model complex interactions. Graph-based approaches mitigate this but struggle with sparsely connected nodes, reducing recommendation accuracy.

\begin{figure*}[ht!]
  \centering
  \includegraphics[width=1.0\linewidth]{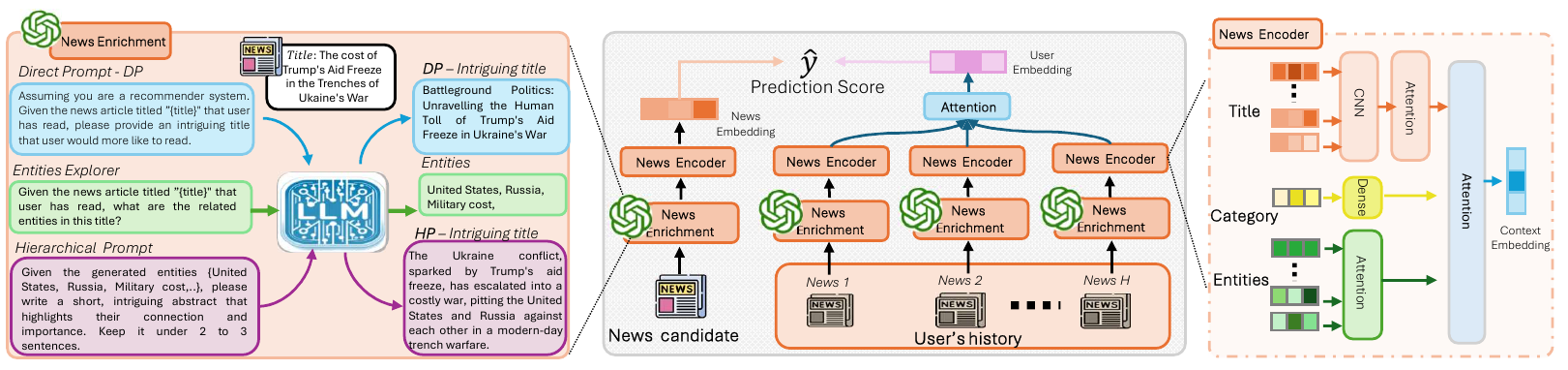}
  \vspace{-0.2in}
  \caption{Illustration of our \myname.}
  \vspace{-0.2in}
  \label{fig:pipeline}
\end{figure*}

% Pre-trained language models (PLMs) have significantly advanced NLP-based recommendation tasks by enhancing text representations \cite{wang2023rethinking}. Models like BERT~\cite{PLM_ENR, UNBERT} improve news and user modeling, while methods such as EmbSum and LLM-TRSR~\cite{EmbSum, LLM-TRSR} summarize user reading histories for refined recommendations, though some summarization techniques may overlook key semantic details.  
% Recent works further leverage Large Language Models (LLMs) to enhance recommendation quality. PLM-NR \cite{PLM_ENR} and UNBERT~\cite{UNBERT} integrate LLM-based embeddings, while LKPNR~\cite{LKPNR} fuses LLMs with Knowledge Graphs (KGs) for better entity alignment. Prompt-based approaches like Prompt4NR~\cite{Prompt4NR} reformulate recommendation tasks as language modeling problems, improving interaction with LLMs. ONCE~\cite{ONCE} employs LLMs as encoders to dynamically summarize user behavior and news content.  
% These advancements demonstrate the potential of LLMs to capture richer contextual information, improving recommendation accuracy and user alignment in news recommendation.

Traditional content-based approaches use natural language processing and machine learning to model user preferences directly from article text. Models such as FIM~\cite{FIM2020}, NAML~\cite{NAML}, and NRMS~\cite{NRMS} employ CNNs, GRUs, and multi-head self-attention to encode news content and user behaviors. More advanced methods like MINER~\cite{miner} and LSTUR~\cite{lstur} enhance user modeling by incorporating poly-attention mechanisms, category-aware weighting, and user ID embeddings for long-term preference capture. To enhance content-based methods, graph-based approaches incorporate external structured knowledge. DKN~\cite{DKN} integrates entity embeddings from knowledge graphs, while DIGAT~\cite{digat} introduces dual-interactive graph attention networks to refine entity-aware recommendations. GLORY ~\cite{GLORY} further refines user representations using graph-based entity extraction. More recently, pre-trained language models (PLMs) and large language models (LLMs) have advanced news recommendation by providing richer text representations and more effective user modeling. PLM-NR~\cite{PLM_ENR}, UNBERT~\cite{UNBERT}, and LLM-TRSR~\cite{LLM-TRSR} leverage PLMs to summarize user reading histories and enhance news encoding, while works such as Prompt4NR~\cite{Prompt4NR} reformulate the recommendation task into a language modeling problem using prompt-based techniques. ONCE~\cite{ONCE} employs LLMs to dynamically summarize both user behavior and news content, demonstrating the potential of LLMs to capture richer contextual and semantic information for improved recommendation quality.

\textbf{Approach.} In this paper, we present our work on leveraging Large Language Models to enhance news recommendation by generating both deeper semantic information and relevant entities from articles, transforming shallow content into richer representations of user preferences. Unlike prior approaches that primarily focus on summarization, our method utilizes LLMs to extrapolate information beyond the title, enriching news articles with contextual details and entity relationships to address data sparsity. By integrating extrapolated entities, we believe that LLM-generated outputs maintain deeper semantic understanding while remaining concise and aligned with the original title. To our knowledge, this is the first study to employ LLMs to generate auxiliary information specifically for news recommendation rather than summarization. We conducted extensive experiments on two benchmark datasets, demonstrating that our approach outperforms recent state-of-the-art methods, effectively capturing richer semantic representations for improved recommendation performance.
\section{METHODOLOGY}
In this section, we will introduce our proposed approach, \myname as illustrated in Fig. 2. A core component of this framework is a news enrichment module comprising hierarchical prompting steps that leverage LLM to generate not only deeper semantic meaning but also concise information from articles, transforming the content into a richer representation.

\subsection{Problem Formulation}
The click history sequence of a user $u$ can be denoted as $H_u = [d_1, d_2,...,d_H]$, where $H$ is the number of historical news articles. Each news article $d_i$ has a title that contains a text sequence $T_i$ consisting of $T$ word tokens and an entity sequence denoted by $E_i$ consisting of $E$ entities. The objective is to predict the probability of interest of a given candidate news article $d_c$ and user $u$.

\subsection{LLM-based News Enrichment}
As demonstrated in several  studies, LLMs can enhance content-based recommendations; however, excessive elaboration on titles may distort their original meaning, reducing the effectiveness of recommendation systems. To address this, a structured prompting approach is crucial to guide LLMs in generating rich and meaningful titles while preserving their intent. Our framework leverages the strengths of LLMs by introducing \textbf{Hierarchical Prompting}, a multi-step reasoning process that systematically refines title generation. This method enables LLMs to process information more efficiently, ensuring the generated titles remain contextually rich, engaging, and aligned with the original content. The workflow follows three key steps: \textbf{(1) Direct Prompting} – Formulate targeted prompts to leverage the capabilities of an LLM in generating an engaging and contextually appropriate title. \textbf{(2) Exploration} – Utilize the LLM’s vast knowledge base to identify and extract relevant entities associated with the article. We believe that if an article has been part of the LLM's training data, it can accurately determine the most relevant title. \textbf{(3) Hierarchical Prompting} – Integrate the initially generated title with the extracted relevant entities, prompting the LLM to refine and generate a more compelling and well-aligned final title. This step is particularly beneficial for news recommendation, as it ensures that the generated title balances informativeness and engagement, preventing misleading or overly generalized headlines while maintaining relevance to user interests. After finish this step, each news article $d_i$ has an updated title that contains a a text sequence ${T'}_i = [w_1, w_2,...,w_{T'}]$ consisting of $T'$ word torkens and a new entity sequence which denoted by ${E'}_i = [e_1, e_2,...,e_{E'}]$ consisting of $E'$ entities.

\subsection{News Encoder}
The news encoder module is designed to learn representations of news articles by incorporating enriched news titles, entities, and topic categories. Similarly to NAML\cite{NAML}, we initialize word embeddings using GloVe and convert a news article's text from a sequence of words $\{w_1, w_2, \dots, w_{T'}\}$ into a sequence of word embedding vectors: $\mathbf{x}_n = [\mathbf{x}_1^\omega, \mathbf{x}_2^\omega, \dots, \mathbf{x}_{T'}^\omega]$. A convolutional neural network (CNN) layer is then applied to learn contextual word representations:

\begin{equation}
    \mathbf{X}_n = \text{CNN}(\mathbf{x}_n) = [\mathbf{x}_1, \mathbf{x}_2, \dots, \mathbf{x}_{T'}].
    \label{eq:Cnn}
\end{equation}

Next, an attention layer is used to identify the most informative words in the titles. Let the attention weight of the \(i^{th}\) word in a news title be denoted as \(\alpha_i^t\), which is computed as follows:

\begin{equation}
    a_i^t = \textbf{q}_t^T \tanh(\mathbf{W}_t \textbf{x}_i^t + \mathbf{b}_t), \quad 
    \alpha_i^t = \frac{\exp(a_i^t)}{\sum_{j=1}^{T} \exp(a_j^t)},
    \label{eq:attention}
\end{equation}
where \(\mathbf{W}_t\) and \(\mathbf{b}_t\) are learnable parameters, and \(q_t\) is the attention query vector. The final news title representation is obtained by computing a weighted sum of contextual word representations: $\mathbf{r} = \sum_{i=1}^{T} \alpha_i^t \textbf{x}_i$. Additionally, we incorporate entity representations to enhance the semantic richness of news embeddings. Fine-grained entity information helps establish connections between news articles when they share common entities, thereby improving the efficiency of the recommendation system. Each news article’s entities are transformed into a sequence of entity embedding vectors: $\mathbf{x}_e = [\mathbf{x}_1^e, \mathbf{x}_2^e, \dots, \mathbf{x}_E^e]$. To learn entity representations \( \mathbf{X}_e \), we employ a self-attention network, followed by an attention mechanism to aggregate multiple entities and derive a unified entity representation, similar to Eq. \ref{eq:attention}. Finally, we merge embedding of category, entity and context \( (\mathbf{X}_c, \mathbf{X}_e, \mathbf{X}_n) \), followed by an additional attention mechanism to obtain the final news representation.

\subsection{User Representation}
The \textbf{user encoder module} derives user representations from the embeddings of their browsed news, as illustrated in Fig \ref{fig:pipeline}. Since different news articles vary in their informativeness, we incorporate a \textbf{news attention network} to prioritize the most relevant items. Given the \textbf{news embeddings} for a user's reading history \( H_u \), an \textbf{attention layer} refines and computes the final \textbf{user embedding}, following a process similar to Eq. \ref{eq:attention}.

% \subsection{Model Training}
% We also use negative sampling techniques to build labeled samples from raw news impression logs, and we use the cross-entropy loss function for model training by classifying which candidate news is clicked. We used Adam as the optimization algorithm and the learning rate was 1e-5. The batch size was 128. For the entity, where we use a pre-trained TransE provided in dataset, with enriched entities and entity for Adressa, we intitialize it randomly.

\section{Experiments}
% \subsection{Datasets}
\textbf{Datasets.} Following \cite{GLORY}, we evaluate our model on two versions of MIND dataset. MIND, sourced from Microsoft News, contains anonymized user behavior logs. The MIND-LARGE version includes one million users with at least five news clicks recorded between October 12 and November 22, 2019, while MIND-SMALL consists of 50,000 users.  
% The Adressa dataset, collected from the Adresseavisen news website, captures user interactions. We use the light version, covering January 1–7, 2017.  
Key dataset statistics are summarized in Table~\ref{tab:datastat}.

\begin{table}[t]
% \vspace{-0.1in}
\caption{Dataset statistics.}
\label{tab:datastat}
\vspace{-0.3cm}
\resizebox{0.8\columnwidth}{!}{%
{\setlength\doublerulesep{0.7pt}
\begin{tabular}{ccrrr}  
\toprule[1pt]%\midrule[0.3pt]
\multirow{2}{*}{Dataset} & \multirow{2}{*}{Version} & \multirow{2}{*}{\# News} & 
\multirow{2}{*}{\# Behaviors} & \multirow{2}{*}{\# Users} \\
\\
% Heading level &  Example & Font size and style\\
\midrule
\multirow{2}{*}{MIND}& SMALL  & 65,238 & 347,727 &  50,000\\
& LARGE  & 161,013 & 24,155,470 & 1,000,000\\
% \hline
% \multirow{1}{*}{ADRESSA}&   & 14,732 & 537,627 &  2,527,571\\

% Charlotte  & 886 & 112,772 &  11,188 & 2,098\\
% Lasvegas   & 868 & 428,782 &  53,304 &  9,995\\
% Phoenix    & 947 & 269,871 & 30,338 &  5,689\\
% Pittsburgh & 905 & 92,449 &  9,116 &  1,710\\
%\midrule[0.3pt] 
\bottomrule[1pt]
\end{tabular}
}%
}
\vspace{-0.3cm}
\end{table}

\textbf{Experimental Setting and Implementation Details.}
% \subsection{Experiment setting and Evaluation}
We use negative sampling (4 negative samples per positive sample) with cross-entropy loss for model optimization. We adopt Adam optimizer \cite{adam} with a learning rate of \( 1 \times 10^{-4} \). We utilize the pre-trained TransE entity representations provided by MIND dataset. Enriched entity representations are initialized randomly. To ensure correctness and prevent redundancy, generated entities are verified using Wikipedia\footnote{https://www.wikidata.org/w/api.php} (e.g., ``United States'' and ``U.S.'' are treated equivalently). We use Gemini 1.0 Pro\footnote{https://huggingface.co/papers/2312.11805} for data enrichment. Following \cite{GLORY}, models are evaluated using AUC, MRR, N@5 (nDCG@5), and N@10 (nDCG@10) over all impressions. %ONCE, GNR, and LKPNR models, which use only MIND subsets, are excluded from direct comparisons.

% \begin{table}[ht!]
%     \centering
%     \caption{Performance of different methods on MIND. There are some missing values denoted as \text{N/a} because we reuse the results from existing works.}
%     \vspace{-0.1in}
%     \label{tab:mind}
% \include{table/full}
%     \vspace{-0.2in}
% \end{table}

\subsection{Empirical Evaluation}

% \subsubsection{Main results}
%\subsubsection{\textbf{Comparison against Baselines.}} 
\textbf{Comparison against Baselines.} We evaluate the performance of various methods on MIND dataset. Our comparison includes conventional news recommendation models, including NAML \cite{NAML}, NPA \cite{npa}, LSTUR \cite{lstur}, and NRMS \cite{NRMS}, alongside recent models, including Glory \cite{GLORY}, DIGAT \cite{digat}, and GLoCIM \cite{glocim}. As shown in Table~\ref{tab:mind}, our model performs the best on MIND-SMALL and second best on MIND-LARGE. Incorporating enriched titles consistently improves the performance of our model. This improvement stems from the enhanced title representations, which better capture semantic connections between similar articles, benefiting text-based news encoders. Additionally, entity representations provide fine-grained information, further boosting recommendation accuracy.

\begin{table}[t]
	\centering
	\caption{Performance on MIND. We reuse the results from existing works, and MINER and DIGAT do not have results on MIND-LARGE. $^{\dagger}$ denotes the second best model.}
	\vspace{-0.3cm}
	\resizebox{\linewidth}{!}{
		\begin{tabular}{c|ccc|ccc}
			\toprule
                \multirow{2}{*}{Model} & \multicolumn{3}{c|}{MIND-SMALL} & \multicolumn{3}{c}{MIND-LARGE} \\
			\cline{2-7}
			{} & AUC & MRR & N@5 (N@10) & AUC & MRR & N@5 (N@10) \\
			\hline
                NPA & 64.65 & 30.01 & 33.14 (39.47) & 65.92 & 32.07 & 34.72 (40.37) \\
                NRMS & 65.63 & 30.96 & 34.13 (40.52) & 67.66 & 33.25 & 36.28 (41.98) \\
                NAML & 66.12 & 31.53 & 34.88 (41.09) & 66.46 & 32.75 & 35.66 (41.40) \\
                LSTUR & 65.87 & 30.78 & 35.15 (40.15) & 67.08 & 32.36 & 35.15 (40.93) \\
                MINER & 68.07 & 32.93 & N.A (42.62) & N.A. & N.A. & N.A. (N.A.) \\ 
                \hline
                DKN & 62.90 & 28.37 & 30.99 (37.41) & 64.07 & 30.42 & 32.92 (38.66) \\
                DIGAT & 67.82 & 32.65 & 36.25 (42.49) & N.A. & N.A. & N.A. (N.A.) \\
                GERL & 65.27 & 30.10 & 32.93 (39.48) & 68.10 & 33.41 & 36.34 (42.03) \\
                GLORY & 67.68 & 32.45 & 35.78 (42.10) & 69.04 & 33.83 & 37.53 (43.69) \\ 
                GLoCIM & 68.21$^{\dagger}$ & 33.02$^{\dagger}$ & 36.69$^{\dagger}$ (42.78$^{\dagger}$) & \textbf{69.52} & \textbf{34.34} & \textbf{37.89} (\textbf{44.08}) \\
                \hline
                \myname (ours) & \textbf{68.88} & \textbf{33.58} & \textbf{37.50 (43.49)} & 69.32$^{\dagger}$ & 34.26$^{\dagger}$ &  37.81$^{\dagger}$ (43.79$^{\dagger}$) \\
			\bottomrule
		\end{tabular}
	}
	\vspace{-0.2cm}
	\label{tab:mind}
\end{table}

\begin{figure}[t]%[h!]
    \centering
    \begin{subfigure}{0.22\textwidth}
        \centering     \includegraphics[width=1\linewidth]{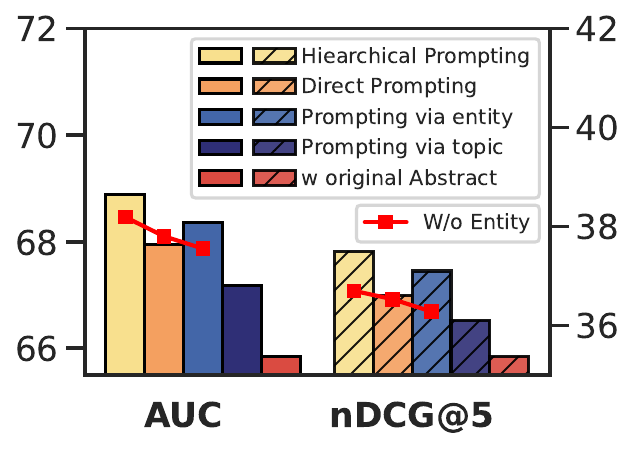}
        % \caption{Different types of prompting}
        \caption{Effect of different promptings on model performance.}
        
    \end{subfigure}
    \begin{subfigure}{0.22\textwidth}
        \centering
        \includegraphics[width=1\linewidth]{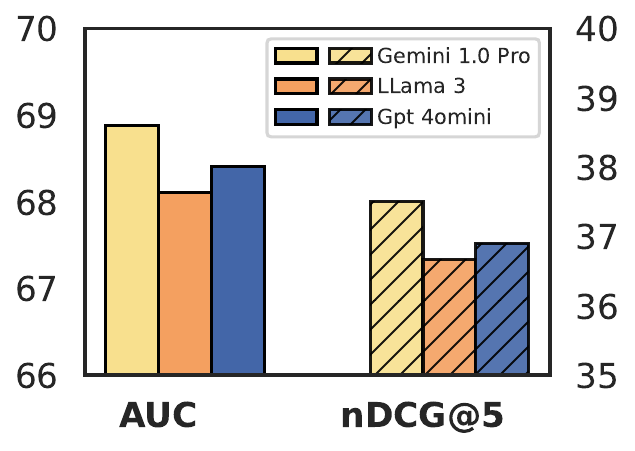}
        % \caption{{Different types ofLLMs}}
        \caption{Effect of different LLMs for data enrichment.}
    \end{subfigure}   
    \vspace{-0.3cm}
    % \caption{Performance of \myname across various aspects on Mind-small.}
    \caption{Effect of different (a) promptings and (b) LLMs.}
    \label{fig:prompt_type}
\end{figure}

\textbf{Analysis of Different Promptings.} Following this, we compare our hierarchical prompting with different prompting strategies. To highlight the effectiveness of our method, we adopt the approach from GNR~\cite{gao2024generative}, which utilizes topics for news representation. Additionally, we evaluate the abstracts provided in the dataset. Furthermore, we compare title generation without hierarchical steps, including direct prompting and entity-based prompting.  As shown in Fig.~\ref{fig:prompt_type}a, hierarchical prompting demonstrates the best performance, followed by direct prompting and entity-based prompting, indicating that structured, stepwise processing enhances recommendation accuracy. Topic-based prompting and using the original abstract show relatively lower performance, suggesting that entity enrichment contributes significantly to model effectiveness.

\textbf{Analysis of Different LLMs for Data Enrichment.} The effectiveness of different LLMs is compared in Fig.~\ref{fig:prompt_type}b. Gemini 1.0 Pro achieves the highest scores for both AUC and nDCG@5, outperforming LLaMA 3\footnote{https://huggingface.co/meta-llama/Meta-Llama-3-8B} and GPT-4o mini\footnote{https://platform.openai.com/docs/models\#gpt-4o-mini} . This suggests that most of LLMs could capture semantic information by following prompting techniques, leading to improved recommendation accuracy.

\subsection{Ablation Study}

%Here we conduct ablation study to better understand our model.

\subsubsection{\textbf{Effect of Pre-trained Language Models for Text Encoding.}} %We acknowledge the effectiveness of pre-trained language models in enhancing news recommendation by improving textual representation learning, as shown in Table~\ref{tab:bert}. 

In this ablation study, we replace our CNN-based text encoder with DistilBERT model. The results in Table~\ref{tab:bert} demonstrate that our approach with a PLM-based encoder consistently outperforms other methods, further validating its effectiveness. This improvement can be attributed to the fact that the original news titles are often sparse and lack sufficient contextual information, making it difficult for PLMs to fully capture user preferences. However, with the enriched and extrapolated titles generated by our method, the input becomes more concise, meaningful, and semantically informative, allowing the PLM to better understand and encode user interests, ultimately leading to enhanced recommendation performance.
\begin{table}[t]%[ht!]
    \centering
    \caption{Performance of different methods on MIND-SMALL using PLM as a new encoder.}
    \vspace{-0.1in}
    \label{tab:bert}
\resizebox{1\columnwidth}{!}{%
{\setlength\doublerulesep{0.7pt}
\renewcommand{\arraystretch}{1.5} % Default value: 1
\begin{tabular}{l|cccccc|c}
\toprule[1pt]%\midrule[0.3pt]
%\hline
          Metrics & NPA & NRMS & LSTUR & CAUM & DIGAT & MINER & \myname\ w/ DistillBert \\ \hline
AUC         & 67.78  & 68.60  & 68.60  & 70.04  & 68.77  & 69.61  & \textbf{70.54}\\
MRR         & 33.24  & 32.97  & 32.97  & 34.71  & 33.46  & 33.97  & \textbf{35.26}\\
N@5         & 36.19  & 36.55  & 36.55  & 37.89  & 37.14  & 37.62  & \textbf{38.07}\\
N@10        & 41.95  & 42.78  & 42.78  & 43.57  & 43.39  & 43.90  & \textbf{44.23}\\ %\hline
%\midrule[0.3pt]
\bottomrule[1pt]
\end{tabular}
}%
}

    \vspace{-0.2in}
\end{table}

\begin{table}[t]%[h!]
    \centering
    \caption{Performance of existing approaches using our enriched title on MIND-SMALL, denoted with ``++''.}
    \label{tab:enrich}
\resizebox{\columnwidth}{!}{%
{\setlength\doublerulesep{0.7pt}
\renewcommand{\arraystretch}{1.5} % Default value: 1
\begin{tabular}{l|cc|cc|cc}
\toprule[1pt]%\midrule[0.3pt]
%\hline
        Metrics & NRMS & NRMS++ & NAML & NAML++ & LSTUR & LSTUR++\\ \hline
AUC         & 65.63  & \textbf{67.86}   & 66.12  & \textbf{68.01}   & 65.87  & \textbf{66.98}   \\
MRR         & 30.96  & \textbf{32.61}   & 31.53  & \textbf{32.75}   & 30.78  & \textbf{32.16}   \\
N@5         & 34.13  & \textbf{36.02}   & 34.88  & \textbf{36.29}   & 35.15  & \textbf{35.82}   \\
N@10        & 40.52  & \textbf{42.26}   & 41.09  & \textbf{42.46}   & 40.15  & \textbf{41.41}   \\ 
%\midrule[0.3pt]
\bottomrule[1pt]
\end{tabular}
}%
}

% NRMS {'auc': np.float64(0.6785714918257374), 'mrr': np.float64(0.32611450372203965), 'ndcg5': np.float64(0.3602652497890186), 'ndcg10': np.float64(0.42262641320905747)}
% NAML {'auc': np.float64(0.6800867459310441), 'mrr': np.float64(0.3274774236007599), 'ndcg5': np.float64(0.3628853823169093), 'ndcg10': np.float64(0.42457987912328476)}

% LSTUR {'auc': np.float64(0.6697971480457978), 'mrr': np.float64(0.32161147858842136), 'ndcg5': np.float64(0.35817500738096743), 'ndcg10': np.float64(0.4141278323467635)}
\vspace{-0.2in}
\end{table}

\subsubsection{\textbf{Performance of Baseline Models Using Our Enriched Title.}} The title generated from our hierarchical prompting further improves conventional baseline models. In this ablation study, we replace original titles with our generated titles and evaluate baseline models. The results in Table~\ref{tab:enrich} confirm the effectiveness of enriched titles in enhancing news recommendations.

\subsubsection{\textbf{Other ablation study.}} 
\begin{figure}[t]%[h!]
    \centering
    
    \begin{subfigure}{0.22\textwidth}
        \centering
        \includegraphics[width=1\linewidth]{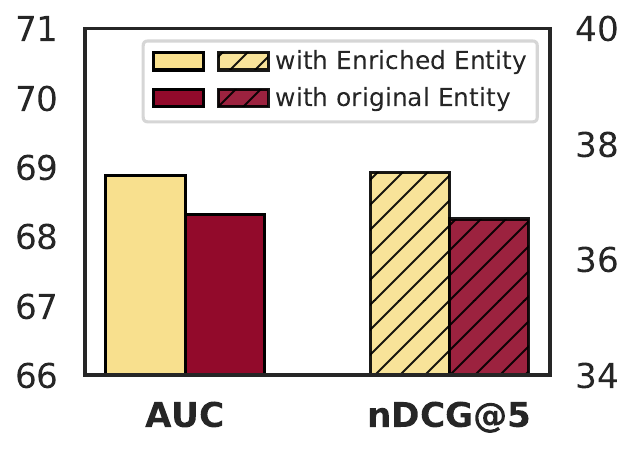}
        \caption{Effect of entity enrichment.}
    \end{subfigure}  
    \begin{subfigure}{0.23\textwidth}
        \centering     \includegraphics[width=1\linewidth]{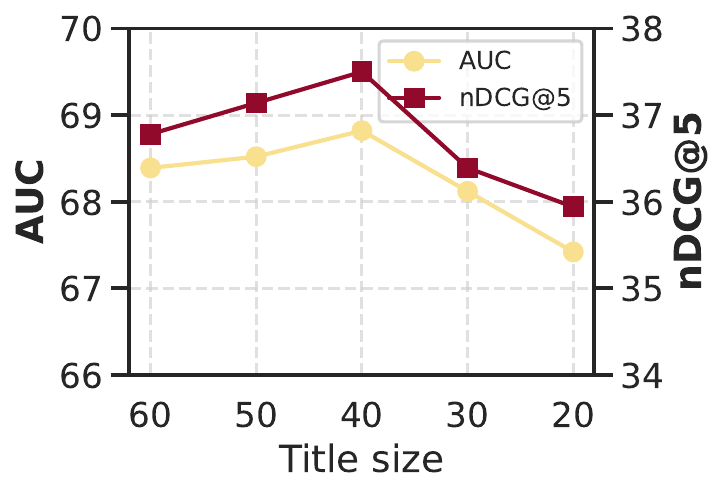}
        \caption{Effect of title token size.}
        
    \end{subfigure}
    \vspace{-0.1in}
    \caption{Performance comparison of \myname with entities provided by MIND-SMALL and our method.}
    \label{EntityEffect}
    \vspace{-0.2in}
\end{figure}

We also compare the enriched entities generated by our approach with the original entities provided in the MIND dataset. As shown in Fig.~\ref{EntityEffect}, incorporating these enriched entities further enhances recommendation performance, demonstrating the value of entity extrapolation. In addition, since the generated titles tend to be longer than the original ones, we conduct an analysis to determine the optimal token length for the new titles. Our results indicate that setting the token length to 40 achieves the best performance.

\section{Conclusion}
In this work, we enhance news recommendation by incorporating enriched titles by utilizing a hierarchical prompting strategy. Experiments demonstrate that these enrichments consistently improve performance across various baseline models. Additionally, PLM-based encoders further boost effectiveness. The results highlight the potential of leveraging LLMs for enriching textual and entity information. %offering a promising direction for future research in news recommendation.
One future work is to connect the generated entities with a knowledge graph to improve recommendation accuracy.

%\newpage

\bibliographystyle{ACM-Reference-Format}
\bibliography{main}

\end{document}